\documentclass[12pt]{article}
\title{On harmonic binomial series}
\author{Mark W. Coffey\\
Department of Physics\\
Colorado School of Mines\\
Golden, CO  80401\\
(Received $\mbox{~~~~~~~~~~~~~~~~~~~~~~~~~~~~~~~2008}$)}
\date{April 29, 2008}
\begin{document}
\maketitle
\baselineskip=25 pt
\begin{abstract}

We evaluate binomial series with harmonic number coefficients, providing
recursion relations, integral representations, and several examples.  The
results are of interest to analytic number theory, the analysis of algorithms,
and calculations of theoretical physics, as well as other applications.

\end{abstract}
 
\vspace{.25cm}
\baselineskip=15pt
\centerline{\bf Key words and phrases}
\medskip 

\noindent
harmonic number, binomial coefficient, digamma function, polygamma function,
generalized hypergeometric function, generalized harmonic number, Legendre function


\bigskip
\centerline{\bf AMS classification numbers}
05A10, 33C20, 33B15 

\baselineskip=25pt
\pagebreak

\medskip
\centerline{\bf Introduction}
\medskip

The evaluation of harmonic number sums has been useful in analytic number theory
for some time (e.g., \cite{berndt1,berndt}).  Recently, the evaluation
of Euler sums \cite{cofcam03,flaj} has been important in various areas
of theoretical physics, including in support of Feynman diagram calculations
(e.g., \cite{cofjcam}).  More recently, the performance of generalized harmonic
number sums has been useful in evaluating Feynman diagram contributions of
perturbative quantum field theory \cite{cof08}.  In addition, harmonic number
sums often arise in the analysis of algorithms.  This especially applies to
algorithms involving searching, sorting, or permutations (e.g., \cite{panprod,zave}).

Here we are interested to evaluate sums containing simultaneously two types of 
special numbers of enumerative combinatorics:  binomial coefficients and
harmonic numbers.  Our methods complement those of Refs. \cite{chu,paule}.
The analytic approach of \cite{chu} is based upon hypergeometric summation
and relies on identities of generalized hypergeometric functions $_{p+1}F_p$ at the
very special argument of $1$.  These identities include the Chu-Vandermonde formula
for $_2F_1$, the Pfaff-Saalsch\"{u}tz theorem for $_3F_2$, the Dougall-Dixon
theorem for $_5F_4$, and the Whipple transformation for $_7F_6$.
In Ref. \cite{paule}, symbolic computation using the Newton-Andrews-Zeilberger
algorithm was used to discover harmonic number relations.  These relations were
motivated by considerations to prove certain 'supercongruences' for Ap\'{e}ry
numbers.

Our results not only complement those of Ref. \cite{chu,paule}, but 
demonstrate that various generalizations are possible, including the 
introduction of a summation parameter.  The latter feature means that many
previous results may be considered special cases, and points out that many
more results should be achievable in the future.

After first introducing some notation, we show how various recursion 
relations for the sums of interest may be developed directly.  We then 
illustrate the development of integral representations for these sums, and
provide examples.  To emphasize the usefulness of the integral representations,
we present low order cases explicitly.  In addition, we give a representation of
binomial-harmonic number sums in terms of the generalized hypergeometric function
and its derivative.  In particular, with regard to the Gauss hypergeometric function,
we are then able to develop results in terms of Legendre function $P_\nu$.

In the final section, we introduce a variation.  Namely,
we consider binomial harmonic sums over the order of the generalized harmonic
numbers.  Additional considerations in analytic number theory motivate the
study of these and related sums \cite{coffeympag}.  In particular, a certain
combination of these sums represents a truncation of a dominating sum for the
Li/Keiper constants. 

We put $H_n \equiv \sum_{k=1}^n 1/k$ for the usual harmonic numbers, and
$H_n^{(r)}$ for the generalized harmonic numbers
$$H_n^{(r)} \equiv \sum_{j=1}^n {1 \over j^r}, ~~~~~~~~~~~~H_n \equiv H_n^{(1)}.  \eqno(0.1)$$
These are given in terms of polygamma functions $\psi^{(j)}$ as
$$H_n^{(r)}={{(-1)^{r-1}} \over {(r-1)!}}\left[\psi^{(r-1)}(n+1)-\psi^{(r-1)}
(1) \right ], \eqno(0.2)$$
where $\psi^{(r-1)}(1)=(-1)^r(r-1)!\zeta(r)$ and $\zeta$ is the Riemann zeta function.
The asymptotic form of $H_n$ is well known, $H_n=\ln n + \gamma +o(1)$, where
$\gamma$ is the Euler constant.  Indeed, by Euler-Maclaurin summation we have
$$H_n=\ln n + \gamma +{1 \over {2n}}+\int_n^\infty {{P_1(x)} \over x^2} dx,
\eqno(0.3)$$
where the periodic Bernoulli polynomial $P_1(x) \equiv B_1(x-[x])=x-[x]-1/2$.
The asymptotic form of $H_n^{(r)}$ for large $n$ is immediately known from that
of $\psi^{(r-1)}(n+1)$, and we have (e.g., \cite{nbs}, p. 260)
$$\psi^{(n)}(z)=(-1)^{n-1}\left[{{(n-1)!} \over z^n} + {{n!} \over {2 z^{n+1}}}
+ O\left({1 \over z^{n+2}}\right)\right].  \eqno(0.4)$$

We investigate here a subclass of generalized harmonic number sums
$$S_n^{(p)}(q,r,m,z) \equiv \sum_{j=0}^n j^p [H_j^{(q)}]^m {n \choose j}^r z^j,  
~~~~~~|z| \leq 1.  \eqno(0.5)$$
In particular, in this paper we restrict for the most part to $m=1$.  When, in addition, $q=r=1$, we write
$$S_n^{(p)}(z) \equiv \sum_{j=0}^n j^p H_j {n \choose j} z^j,  
~~~~~~|z| \leq 1.  \eqno(0.6)$$
When simply $p=0$, we omit the superscript.

We note a very convenient recursion for the general sums of Eq. (0.5):
$$S_n^{(p+1)}(q,r,m,z)=z {\partial \over {\partial z}}S_n^{(p)}(q,r,m,z).  
\eqno(0.7)$$
In this way, from an initial sum, successive sums may be obtained.

Although we do not follow this line of inquiry, we mention that many integral
representations for binomial coefficients are known.  These (e.g., \cite{grad},
pp. 372-375; \cite{kaucky}, p. 53) include \cite{grad}, (p. 375)
$${n \choose m}={2^{n+2} \over \pi}\int_0^{\pi/2} \cos^n x \sin nx \sin 2mx ~dx
={2^{n+2} \over \pi}\int_0^{\pi/2} \cos^n x \cos nx \cos 2mx ~dx.  \eqno(0.8)$$
As a contour integral, we have for complex $\alpha$ and integral $j \geq 1$,
$${\alpha \choose j}={1 \over {2\pi i}}\int_{|z|=r} (1+z)^\alpha z^{-j-1} dz,
~~~~~~0 < r < 1.  \eqno(0.9)$$
We may note that this equation is particularly well formed for taking derivatives
with respect to the parameter $\alpha$.
If both $n \geq 1$ and $j \geq 1$ are integral, we no longer have a branch point
at $z=-1$, and may write
$${n \choose j}={1 \over {2\pi i}}\int_{|z|=r} (1+z)^n z^{-j-1} dz,
~~~~~~0 < r < \infty.  \eqno(0.10)$$
Equations (0.9) and (0.10) may be immediately verified by using the binomial
theorem to compute the residue of the integrand at $z=0$.


\medskip
\centerline{\bf Recursion relations for binomial-harmonic number sums}
\medskip

We have 
\newline{\bf Proposition 1}.  We have the recursion relation (a)
$$S_{n+1}(1)=2S_n(1) + {{2^{n+1}-1} \over {n+1}}, \eqno(1.1)$$
and (b)
$$S_{n+1}(z)=(1+z)S_n(z) + {{(z+1)^{n+1}-1} \over {n+1}}. \eqno(1.2)$$
For part (a), $S_1(1)=1$, and for part (b), $S_1(z)=z$.

{\bf Proposition 2}.  We have the recursion relation (a)
$$S_{n+1}^{(p)}(1)=2S_n^{(p)}(1)+\sum_{\ell=0}^{p-1} {p \choose \ell} S_n^{(\ell)}
+{n \over 2} \sum_{\ell=0}^p {p \choose \ell}  ~ _{\ell+1}F_\ell(2,2,\ldots,2,1-n;1,1,\ldots,1,3;-1),
\eqno(1.3)$$
and (b)
$$S_{n+1}^{(p)}(z)=(1+z)S_n^{(p)}(z)+z\sum_{\ell=0}^{p-1} {p \choose \ell} S_n^{(\ell)} +{n \over 2} z^2 \sum_{\ell=0}^p {p \choose \ell}  ~_{\ell+1}F_\ell(2,2,\ldots,2,1-n;1,1,\ldots,1,3;-z).  \eqno(1.4)$$
For part (a), $S_1^{(p)}(1)=1$, and for part (b), $S_1^{(p)}(z)=z$.

{\bf Proposition 3}.  We have the recursion relation (a)
$$S_n^{(1)}(z)=nzS_{n-1}(z)+(1+z)^n-1.  \eqno(1.5)$$
Let 
$$\beta_n(p,z) \equiv \sum_{j=1}^n j^{p-1}{{n-1} \choose {j-1}} {z^j \over j}
=z ~_{p-1}F_{p-2}(2,2,\ldots,2,1-n;1,1,\ldots,1;-z).  \eqno(1.6)$$
Then we have (b)
$$S_n^{(p)}(z)=nz \sum_{\ell=0}^{p-1} {{p-1} \choose \ell} S_{n-1}^{(\ell)}(z) +n\beta_n(p,z).  \eqno(1.7)$$

%
%

{\it Proof} of Proposition 1.  The proof of part (b) subsumes that of part (a).
By using a recursion relation for the binomial coefficient \cite{nbs} (p. 822),
we have
$$S_{n+1}(z)= \sum_{j=0}^{n+1} H_j {{n+1} \choose j} z^j=
\sum_{j=0}^{n+1} H_j \left[{n \choose j} +{n \choose {j-1}}\right] z^j$$
$$=S_n(z) +\sum_{j=1}^{n+1} H_j {n \choose {j-1}}z^j, \eqno(1.8)$$
where we have used the facts ${n \choose {n+1}}=0=H_0$.  Further, by shifting
the summation index and using the recursion relation for harmonic numbers, we 
have
$$S_{n+1}(z)= S_n(z)+\sum_{j=0}^n H_{j+1} {n \choose j} z^{j+1}$$
$$=S_n(z)+\sum_{j=0}^n \left(H_j +{1 \over {j+1}}\right){n \choose j} z^{j+1}$$
$$=(1+z)S_n(z)+ {{(z+1)^{n+1}-1} \over {n+1}}, \eqno(1.9)$$
wherein the latter sum may be obtained by integrating the binomial theorem.

{\it Proof} of Proposition 2a.  For this part, omitting the $z=1$ argument, we 
have
$$S_{n+1}^{(p)}= \sum_{j=0}^{n+1} j^p H_j {{n+1} \choose j}=
\sum_{j=0}^{n+1} j^p H_j \left[{n \choose j} +{n \choose {j-1}}\right]$$
$$=S_n^{(p)} +\sum_{j=1}^{n+1} j^p H_j {n \choose {j-1}}$$
$$=S_n^{(p)}+ \sum_{j=0}^n (j+1)^p H_{j+1} {n \choose j}$$
$$=S_n^{(p)}+ \sum_{j=0}^n \sum_{\ell=0}^p {p \choose \ell}j^\ell
\left(H_j +{1 \over {j+1}}\right){n \choose j}.  \eqno(1.10)$$
We next interchange the order of the two sums, and to complete this part of
the Proposition we need to perform the sum
$$\sum_{j=1}^n {j^\ell \over {(j+1)}} {n \choose j}=\sum_{j=0}^{n-1} \left[
{{(2)_j} \over {(1)_j}}\right]^\ell {1 \over {j+2}} {n \choose {j+1}},  
\eqno(1.11)$$
where $(a)_j=\Gamma(a+j)/\Gamma(a)$ is the Pochhammer symbol, and $\Gamma$ is the
Gamma function.  We further use $(2)_j/(3)_j=2/(j+2)$ and 
$${n \choose {j+1}}= {{(-1)^{j+1}} \over {(j+1)!}} (-n)_{j+1}={{(-1)^j} \over {(j+1)!}} n (1-n)_j . \eqno(1.12)$$
Appealing to the series definition of $_{\ell+1}F_\ell$ completes part (a).
The steps for part (b) are very similar, and are omitted.
 
{\it Proof} of Proposition 3a. We use the property $j{n \choose j}=n {{n-1}
\choose {j-1}}$, so that
$$S_n^{(1)}(z)=n\sum_{j=1}^n {{n-1} \choose {j-1}}\left({1 \over j}+H_{j-1}\right
)z^j$$
$$=n\left[\sum_{j=1}^{n-1} {{n-1} \choose {j-1}}H_{j-1}z^j +H_{n-1}z^n+
\sum_{j=1}^n {{n-1} \choose {j-1}} {z^j \over j}\right]$$
$$=n\left[\sum_{j=1}^{n-2} {{n-1} \choose j}H_jz^{j+1} +H_{n-1}z^n+
\sum_{j=1}^n {{n-1} \choose {j-1}} {z^j \over j}\right]$$
$$=n\left[\sum_{j=1}^{n-1} {{n-1} \choose j}H_{j-1}z^{j+1}-H_{n-1}z^n +H_{n-1}z^n+
\sum_{j=1}^n {{n-1} \choose {j-1}} {z^j \over j}\right]$$
$$=n\left[zS_{n-1}(z)+{1 \over n}\left((1+z)^n-1\right)\right], \eqno(1.13)$$
wherein the latter sum may be obtained by integrating the binomial theorem.

For part (b), we proceed similarly,
$$S_n^{(p)}(z)=\sum_{j=1}^n j^{p-1} j{n \choose j}H_j z^j$$
$$=n\sum_{j=1}^n j^{p-1}{{n-1} \choose {j-1}}H_jz^j$$
$$=n\left[\sum_{j=1}^{n-1} {{n-1} \choose {j-1}}H_jz^j +n^{p-1} H_nz^n\right]$$
$$=n\left[\sum_{j=1}^n j^{p-1}{{n-1} \choose {j-1}}\left({1 \over j}+H_{j-1}\right
)z^j+n^{p-1}H_nz^n\right]$$
$$=n\left[\beta_n(p,z)+\sum_{j=0}^{n-1} (j+1)^{p-1} {{n-1} \choose j} H_j z^{j+1}
\right].  \eqno(1.14)$$
We then binomially expand, interchange sums, and the Proposition is complete.

{\it Remark}.  Similar considerations can be given for quadratic and more generally
nonlinear sums, that are outside of the present investigation.

\medskip
\centerline{\bf Examples}  
\medskip

The recursion relations (1.1) and (1.2) with initial condition may be explicitly solved.  For Eq. (1.1) we have
$$S_n(1)=2^n H_n +{1 \over {n+1}}~_2F_1(1,1;n+2;-1) -2^n \ln 2.  \eqno(1.15)$$
Since easily we have
$${1 \over {n+1}}~_2F_1(1,1,n+2;-1) -2^n \ln 2=-2^n \sum_{j=1}^n {1 \over {j2^j}},
\eqno(1.16)$$
we obtain
$$S_n(1)=2^n\left(H_n-\sum_{j=1}^n {1 \over {j2^j}}\right),  \eqno(1.17)$$
a known result \cite{paule}.  For Eq. (1.2) we find
$$S_n(z)=(1+z)^n\left[H_n+{1 \over {(z+1)^{n+1}}}{1 \over {(n+1)}} ~_2F_1\left(1,n+1;n+2;{1 \over {z+1}}\right) +\ln \left({z \over {z+1}}\right)\right].  \eqno(1.18)$$
By using a transformation formula \cite{grad} (p. 1043), we may write this as 
$$S_n(z)=(1+z)^n\left[H_n+{1 \over {z(z+1)^n}}{1 \over {(n+1)}} ~_2F_1\left(1,1;n+2;-{1 \over z}\right) +\ln \left({z \over {z+1}}\right)\right].  \eqno(1.19)$$
For $z=1$ here we have the explicit reduction to Eqs. (1.15) and (1.17).  At
$z=-1$ we have
$$~_2F_1\left(1,1;n+2;1\right) ={{\Gamma(n+2)\Gamma(n)} \over {\Gamma^2(n+1)}}
={{n+1} \over n}.  \eqno(1.20)$$
Therefore, we obtain $S_n(-1)=-1/n$.

From Eqs. (1.5) and (1.18) we obtain
$$S_n^{(1)}(z)=nz(1+z)^{n-1}\left[H_{n-1}+{1 \over {(z+1)^n}}{1 \over n} ~_2F_1\left(1,n;n+1;{1 \over {z+1}}\right) +\ln \left({z \over {z+1}}\right)\right].  \eqno(1.21)$$

{\it Remark}.  From the recursion relations of Proposition 2, it appears that the
portions of the respective binomial-harmonic series containing harmonic numbers
are given for $p \geq 1$ by
$$\tilde{S}_n^{(p)}(1)=2^{n-p} (n)_p \left[H_{n-p}-\sum_{j=1}^{n-p} {1 \over {j2^j}}
\right], \eqno(1.22)$$
and
$$\tilde{S}_n^{(p)}(z)=(1+z)^{n-p} (n)_p \left[H_{n-p}+{1 \over {z(z+1)^{n-p}}}
{1 \over {(n+1-p)}} ~_2F_1\left(1,1;n-p+2;-{1 \over z}\right) -\ln \left({{z+1} \over z}\right)\right]. \eqno(1.23)$$

\medskip
\centerline{\bf Integral representations for linear binomial-harmonic number sums}  
\medskip

We have 
\newline{\bf Proposition 4}.  We have the integral representation (a)
$$S_n^{(p)}(z)=nz\int_0^1 \left[t ~_pF_{p-1}(2,2,\ldots,2,1-n;1,1,\ldots,1;-zt)\right.$$
$$\left. -~_pF_{p-1}(2,2,\ldots,2,1-n;1,1,\ldots,1;-z)\right]{{dt} \over {t-1}}, \eqno(2.1)$$
and (b)
$$S_n^{(p)}(q,1,1,z)={{(-1)^q} \over {(q-1)!}}nz\int_0^1 \left[t ~_pF_{p-1}(2,2,\ldots,2,1-n;1,1,\ldots,1;-zt)\right. $$
$$\left. - ~_pF_{p-1}(2,2,\ldots,2,1-n;1,1,\ldots,1;-z)\right]{{\ln^{q-1} t} \over {t-1}}dt . \eqno(2.2)$$

{\it Proof}.  For part (a), we employ the relation $H_j=\psi(j+1)+\gamma$, where
$\psi=\Gamma'/\Gamma$ is the digamma function, together with an integral
representation for this function \cite{grad} (p. 943):
$$S_n^{(p)}(z)=\sum_{j=0}^n j^p [\psi(j+1)+\gamma]{n \choose j} z^j$$
$$=\sum_{j=0}^n j^p {n \choose j} z^j \int_0^1 {{t^j-1} \over {t-1}} dt .
\eqno(2.3)$$
The integral is absolutely convergent and we may interchange summation and
integration.  We then rewrite the summation, using relation (1.12), and apply the
series definition of $_pF_{p-1}$.  Equation (2.1) follows.

For part (b), we use Eq. (0.2), writing
$$S_n^{(p)}(q,1,1z)={{(-1)^q} \over {(q-1)!}}\sum_{j=0}^n j^p [\psi^{(q-1)}(j+1) -\psi^{(q-1)}(1)]{n \choose j} z^j$$
$$={{(-1)^q} \over {(q-1)!}}\sum_{j=0}^n j^p {n \choose j} z^j \int_0^1 \left(
{{t^j-1} \over {t-1}}\right ) \ln^{q-1} t ~dt . \eqno(2.4)$$
Here, we employed the result of multiply differentiating an integral representation
for the digamma function to obtain that for the polygamma function.  The integral
is again absolutely convergent and we may interchange it with the summation.
Carrying out the summation, we have Eq. (2.2).

{\it Remarks}.  The integral representation of the polygamma function used above,
$$\psi^{(q-1)}(z)=\int_0^1 \left({{t^{z-1}-1} \over {t-1}}\right ) \ln^{q-1} t ~dt
=(-1)^q (q-1)! \zeta(q,z), \eqno(2.5)$$
shows the connection with the Hurwitz zeta function $\zeta(s,a)$ at integer values
of $q$.

Reference \cite{cof08} contains a number of other integral representations
for harmonic numbers.  Similarly, these may be applied to obtain other integral
representations of the sums $S_n^{(p)}$.

The direct verification of the general property (0.7) from the integral
representations requires the result
$${\partial \over {\partial z}} ~_pF_{p-1}(2,2,\ldots,2,1-n;1,1,\ldots,1;-zt)$$
$$=(n-1)2^{p-1} t ~_pF_{p-1}(3,3,\ldots,3,2-n;2,2,\ldots,2;-zt), \eqno(2.6)$$
as well as a transformation formula for $_pF_{p-1}$.  It is possible that this
transformation may be effected by noting 
$$\left[{{(3)_j} \over {(2)_j}}\right]^{p-1}=\left(1+{j \over 2}\right)^{p-1}
={1 \over 2^{p-1}}\left[{{(2)_j} \over {(1)_j}}+1\right]^{p-1}, ~~~~
(2-n)_j=\left(1+{j \over {1-n}}\right)(1-n)_j .  \eqno(2.7)$$

To show the utility of integral representations for binomial-harmonic number
sums, we specialize in the following two sections to low order cases of the sums $S_n^{(p)}$, and provide corresponding details.

\medskip
\centerline{\bf Explicit expressions for sums $S_n^{(p)}(z)$}  
\medskip

We have
{\newline \bf Proposition 5}.  We have for $|z| \leq 1$,
$$S_n(z)=n z(1+z)^{n-1} ~_3F_2\left(1,1,1-n;2,2;{z \over {1+z}}\right),  
\eqno(3.1)$$
giving
{\newline \bf Corollary 1}.
$$S_n^{(1)}(z)=(1+z)^{n-2}\left\{1+z-{1 \over {(1+z)^{n-1}}}+ n^2 z^2 ~_3F_2\left(1,1,1-n;2,2;{z \over {1+z}}\right) \right\},  \eqno(3.2a)$$
and
$$S_n^{(2)}(z)=z(1+z)^{n-3}\left\{(1+z)\left[2n-1+{{1-n} \over {(1+z)^n}}\right]
+ n^2 z(1+nz) ~_3F_2\left(1,1,1-n;2,2;{z \over {1+z}}\right) \right\}.
\eqno(3.2b)$$

The right side of Eq. (3.1) is easily verified to be a polynomial of degree
$n$ in $z$, as it must.  For the (terminating) $_3F_2$ function is a polynomial
of degree $n-1$ in $z/(1+z)$.  When multiplied by the prefactor of $z(1+z)^{n-1}$,
we obtain a polynomial of degree $n$.  We could continue Corollary 1 indefinitely,
but these instances serve our current purpose.

{\it Proof}.  As with Eq. (2.3) at $p=0$, we have  
$$S_n(z)=\sum_{j=0}^n {n \choose j} z^j \int_0^1 {{t^j-1} \over {t-1}} dt$$
$$=-\int_0^1 [(1+z)^n-(1+zt)^n] {{dt} \over {t-1}}.  \eqno(3.3)$$
Changing variable with $u=1-t$, we have
$$S_n(z)=(1+z)^n \int_0^1\left\{1-\left[1 - {{zu} \over {1+z}}\right]^n \right\}
{{du} \over u},  \eqno(3.4)$$
and then obtain the Proposition.

Corollary 1 follows by applying property (0.7).

\pagebreak
\centerline{\bf Explicit expressions for some sums $S_n^{(p)}(1)$ in terms of
harmonic numbers}  
\medskip

We have
{\newline \bf Proposition 6}.  We have 
$$S_n(1)=2^n\left[H_n-\sum_{j=1}^n {1 \over {j2^j}}\right], \eqno(4.1a)$$
$$S_n^{(1)}(1)=n 2^{n-1}\left[H_{n-1}-\sum_{j=1}^{n-1} {1 \over {j2^j}}\right]-1+2^n,
\eqno(4.1b)$$
and
$$S_n^{(2)}(1)=2^{n-2}\left\{n(n+1)\left[H_{n-2}-\sum_{j=1}^{n-2}{1 \over {j2^j}}\right] +[2n(1-2^{-n})-1]{{2n} \over {n-1}} \right \}.  \eqno(4.1c)$$

Although Eqs. (4.1a) and (4.1b) are known results \cite{paule}, our method of
proof is different.  In addition, the latter appears to be a rediscovery of
earlier closed form summations \cite{kaucky}.

{\it Proof}.  Using the integral representation of the previous section,
writing Eqs. (3.3) and (3.4) at $z=1$, we have
$$S_n(1)=\int_0^1 [-2^n-(1+t)^n] {{dt} \over {t-1}}$$
$$=2^n \int_0^1\left[1-\left(1 - {u \over 2}\right)^n \right] {{du} \over u}$$
$$=2^n\left[\psi(n+1)+\gamma+\int_{1/2}^1 {{(1-w)^n} \over w} dw-\ln 2\right]$$
$$=2^n\left[H_n+ {1 \over {2^n (n+1)}} ~_2F_1(1,1,n+2;-1) - \ln 2\right].
\eqno(4.2)$$
Here we have applied Eqs. (A5) and (A6) of Ref. \cite{coffeympag}.  By Eq. (1.16),
Eq. (4.1a) follows.

Writing Eq. (2.3) at $z=1$ and $p=1$, we have  
$$S_n^{(1)}(1)=\sum_{j=0}^n j{n \choose j} z^j \int_0^1 {{t^j-1} \over {t-1}} dt$$
$$=n\int_0^1 [t(1+t)^{n-1}-2^{n-1}] {{dt} \over {t-1}}$$
$$=n 2^{n-1}\int_0^1 \left[t\left({{t+1} \over 2}\right)^{n-1}-1\right] {{dt} \over 
{t-1}}.  \eqno(4.3)$$
Changing variable with $u=1-t$, we have
$$S_n^{(1)}(1)=n2^{n-1} \int_0^1\left[1-\left(1 - {u \over 2}\right)^{n-1} \right]
{{du} \over u}-n2^{n-1}\int_0^1 \left(1 - {u \over 2}\right)^{n-1} du$$
$$=n2^{n-1}\left[H_{n-1}-\sum_{j=1}^{n-1} {1 \over {j2^j}}\right]-2^n(2^{-n}-1),  
\eqno(4.4)$$
giving Eq. (4.1b).

Following a similar procedure, writing Eq. (2.3) at $z=1$ and $p=2$, we have  
$$S_n^{(2)}(1)=\sum_{j=0}^n j^2{n \choose j} z^j \int_0^1 {{t^j-1} \over {t-1}} dt$$
$$=n\int_0^1 [nt(1+nt)(1+t)^{n-2}-2^{n-2}n (n+1)] {{dt} \over {t-1}}$$
$$=n 2^{n-2}\int_0^1 \left[nt(1+nt)\left({{t+1} \over 2}\right)^{n-2}-n(n+1)\right] 
{{dt} \over {t-1}}.  \eqno(4.5)$$
Now with $u=1-t$, we have
$$S_n^{(2)}(1)=2^{n-2}\left\{n(1-u)[(n+1)-nu]\left(1-{u \over 2}\right)^{n-2}
-n(n+1)\right\} {{du} \over u}$$
$$=2^{n-2}\left\{n(n+1)\int_0^1 \left[\left(1-{u \over 2}\right)^{n-2}-1\right]{{du}
\over u}
+\int_0^1 \left(1-{u \over 2}\right)^{n-2}[-2n^2-n+n^2u]du \right \}$$
$$=2^{n-2}\left\{n(n+1)\left[H_{n-2}-\sum_{j=1}^{n-2} {1 \over {j2^j}}\right]\right.$$
$$\left. +n(1+2n) {2 \over {n-1}}(2^{1-n}-1)+{{4n} \over {n-1}} [1-2^{-n}(1+n)] \right\}.  \eqno(4.6)$$
Simplifying the latter terms on the right side of this equation gives Eq. (4.1c),
and completes the Proposition.

\medskip
\centerline{\bf Hypergeometric approach}  
\medskip

We have
{\newline \bf Proposition 7}.  For integers $p \geq 1$,
$$S_n^{(0)}\left(1,p,1,{1 \over x}\right)=-{x^{-n} \over p} {\partial \over
{\partial \nu}} ~_pF_{p-1} [-\nu,\ldots,-\nu;1,\ldots,1;(-1)^p x]|_{\nu=n}$$
$$ + H_n x^{-n} ~_pF_{p-1}[-n,\ldots,-n;1,\ldots,1;(-1)^p x].  \eqno(5.1)$$

{\it Proof}.  We have
$${\nu \choose j}=(-1)^j {{(-\nu)_j} \over {(1)_j}}, \eqno(5.2)$$
so that
$$\sum_{j=0}^\infty {\nu \choose j}^p x^j =~_pF_{p-1}
[-n,\ldots,-n;1,\ldots,1;(-1)^p x].  \eqno(5.3)$$
Using
$${\partial \over {\partial \nu}}(-\nu)_j^p=p(-\nu)_j^p [\psi(\nu+1)-\psi(\nu-j+1)],
\eqno(5.4)$$
and the sum
$$\sum_{j=0}^n  {n \choose j}^p H_{n-j}x^j = x^n\sum_{m=0}^n {n \choose m}^p H_m x^{-m} = x^n S_n^{(0)}\left(1,p,1,{1 \over x}\right), \eqno(5.5)$$
we obtain the Proposition.


We focus on the $p=2$ case of Proposition 7, and obtain a number of series
representations.  For this, we introduce the Legendre functions of the first
kind $P_\nu$, and the function
$$R_n(z)=\left.{{\partial P_\nu(z)} \over {\partial \nu}}\right|_{\nu=n}-
\ln\left({{1+z} \over 2}\right) P_n(z), \eqno(5.6)$$
where $n \geq 0$ is an integer, and $P_n$ is a Legendre polynomial.  We have
{\newline \bf Proposition 8}.
$$S_n^{(0)}\left(1,2,1,{1 \over x}\right)=-{x^{-n} \over 2} {\partial \over
{\partial \nu}} ~_2F_1 (-\nu,-\nu;1;x)|_{\nu=n}$$
$$ + H_n x^{-n} ~_2F_1(-n,-n;1;x).$$
$$=-{x^{-n} \over 2} (1-x)^n R_n\left({{1+x} \over {1-x}}\right)+H_n x^{-n}
(1-x)^n P_n\left({{1+x} \over {1-x}}\right).  \eqno(5.7)$$

{\it Proof}.  We first observe that through the use of a transformation
formula for $_2F_1$ we have
$$P_\nu(1-2x)= ~_2F_1(-\nu,\nu+1;1;x)=(1-x)^\nu ~_2F_1\left(-\nu,-\nu;1;{x \over
{x-1}} \right),  \eqno(5.8)$$
giving
$$P_\nu(z)=\left({{z+1} \over 2}\right)^\nu ~_2F_1\left(-\nu,-\nu;1;{{z-1} \over
{z+1}}\right).  \eqno(5.9)$$
Differentiation of this equation with respect to $\nu$ and comparison with
the defining relation (5.6) yields
$$R_n(z)=\left({{z+1} \over 2}\right)^n \left.{\partial \over {\partial \nu}} ~_2F_1\left(-\nu,-\nu;1;{{z-1} \over {z+1}}\right)\right|_{\nu=n}.  \eqno(5.10)$$
With a change of variable, we obtain Eq. (5.7).

We now discuss and illustrate Proposition 8, in the course of which we obtain
{\newline \bf Corollary 2}
$$S_n^{(0)}(1,2,1,1)=\sum_{j=0}^n H_j {n \choose j}^2=(2H_n-H_{2n}){{2n} \choose n},
\eqno(5.11)$$
recovering a known result \cite{paule}.  

There are several series representations of the function $R_n$ available, 
including the Bromwich forms \cite{bromwich}
$$R_n(z)=\sum_{k=1}^n {1 \over k}[P_k(z)-P_{k-1}(z)]P_{n-k}(z), \eqno(5.12)$$
$$R_n(z)=2\sum_{k=0}^{n-1} (-1)^{n+k} {{2k+1} \over {(n-k)(n+k+1)}}[P_k(z)-P_n(z)],
\eqno(5.13)$$
and from a formula of Jolliffe \cite{jolliffe},
$$R_n(z)=-2 \ln\left({{1+z} \over 2}\right) P_n(z)+{1 \over {2^{n-1}n!}}{d^n \over {dz^n}} \left[(z-1)^n(z+1)^n \ln\left({{z+1} \over 2}\right)\right]. \eqno(5.14)$$
Besides these older results, $R_n$ and closely related polynomials have been
very recently restudied \cite{rs}.  
Since the Legendre polynomial has $P_n(1)=1$ and $P_k(-1)=(-1)^k$, we readily
see that the function $R_n$ satisfies $R_n(1)=0$ and $R_n(-1)=2(-1)^n H_n$.  It is
evident from the Bromwich formula (5.13) that
$$R_n(z)=2(H_{2n}-H_n)P_n(z)+2\sum_{k=0}^{n-1} (-1)^{n+k} {{2k+1} \over {(n-k)(n+k+1)}}P_k(z). \eqno(5.15)$$
The property $R_n(-1)=2(-1)^n H_n$ follows immediately from this equation.
With the use of the duplication formula satisfied by the digamma function,
the property $R_n(1)=0$ may also be deduced from Eq. (5.15).

Equivalent to the Chu-Vandermonde summation $_2F_1(-n,-n;1;1)={{2n} \choose n}$,
is the relation, via Eq. (5.9), 
$$\lim_{x \to 1} (1-x)^n P_n\left({{1+x} \over {1-x}}\right)={{2n} \choose n}.  \eqno(5.16)$$
In addition, with the help of Eq. (5.15) and the use of partial fractions, we have
$$\lim_{x \to 1} (1-x)^n R_n\left({{1+x} \over {1-x}}\right)=-2{{2n} \choose n}
(H_{2n}-H_n).  \eqno(5.17)$$
Therefore, from Proposition 8 we obtain Corollary 2.

By the same method of this section, on taking more derivatives with respect to
$\nu$ in Eq. (5.1), it is possible to get higher values of $q$ in the sums
defined in Eq. (0.5).   In this regard, we very briefly mention the functions
$$Q(x,z)=~_2F_1(x,-x;1;z)=\sum_{j=0}^\infty {{(x)_j (-x)_j} \over {(j!)^2}} z^j, ~~~~~~ Q(x) \equiv Q(x,-1), \eqno(5.18)$$
and the expansion
$$Q(x)=1 + \sum_{j=1}^\infty A_{2j} x^{2j}, \eqno(5.19)$$
written in Ref. \cite{rutledge}.  Obviously the Maclaurin coefficients here are
given by
$$A_{2j}= {1 \over {(2j)!}} \left. \left({d \over {dx}}\right)^{2j} Q(x)\right|_{x=0}.
\eqno(5.20)$$
Then by the property (5.4) at $p=1$, the coefficients $A_{2j}$ involve sums and
products of generalized harmonic numbers. Further connections with Legendre 
functions and Stirling numbers of the first kind we do not pursue here.  

Among the generalized hypergeometric functions that reduce to polynomials is the
family
$$H_n(x,a,z)= ~_3F_2(-n,n+1,x;1,a;z), \eqno(5.21)$$
where $n \geq 0$ is an integer and $x$, $a \in C$, with $a\neq -n-1,-n-2, \ldots$
These functions may be obtained by a certain integration over shifted Legendre
polynomials,  
and similarly for higher order functions $_pF_{p-1}$.  
Obviously the set of functions of Eq. (5.21) includes $H_n(n+1,1,z)$ and 
related functions.  One may then ask for a generalization of Proposition 8 for
the case of $p=3$ in Proposition 7, an effort that we are leaving to future work.


We also mention an integral representation for the sum of Proposition 8, and
how the case of Corollary 2 may be otherwise obtained.
We have
$$S_n^{(0)}(1,2,1,z)=\int_0^1 [~_2F_1(-n,-n,1,zt)-~_2F_1(-n,-n,1,z)]{{dt} \over
{t-1}}. \eqno(5.22)$$
As special case, we have
$$S_n^{(0)}(1,2,1,1)={{2n} \choose n}\int_0^1 \left[{1 \over {{{2n} \choose n}}}
~_2F_1(-n,-n,1,t)-1\right]{{dt} \over {t-1}} \eqno(5.23a)$$
$$=-{{2n} \choose n} \int_0^\infty \left[{1 \over {{{2n} \choose n}}}
(1-x)^{-n} P_n(1-2x) -1\right]{{dx} \over {x-1}} \eqno(5.23b)$$
$$=-{{2n} \choose n} \int_1^\infty \left[{1 \over {{{2n} \choose n}}}
\left({2 \over {1+z}}\right)^n P_n(z) -1\right]{{dz} \over {1+z}}, \eqno(5.23c)$$
where we changed variable and used Eqs. (5.8) and (5.9).
One way to demonstrate equality with the result of Corollary 2 is to show that
the integral of this equation satisfies the same recursion relation 
$Q_{n+1}-Q_n=1/(2n+1)+1/(2n+2)-2/(n+1)$ as the quantity $Q_n = H_{2n}-2H_n$,
with initial condition $Q_1=-1/2$.  
We also note the integral representation
$$Q_n=-\int_0^1 {{(t^n-1)^2} \over {t-1}} dt.  \eqno(5.24)$$

\medskip
\centerline{\bf Other class of binomial harmonic sums}  
\medskip

Finally, we describe another class of sums, where the summation is now over the
order of the generalized harmonic numbers,
$${\cal S}_n(M,z) \equiv \sum_{m=2}^n {n \choose m} H_M^{(m)} z^m.  \eqno(6.1)$$
We have
\newline{\bf Proposition 9}.  Let $L_n^\alpha$ be the associated Laguerre 
polynomial of degree $n$ (e.g., \cite{nbs,coffeympag,grad}).  We have
$${\cal S}_n(M,z) =z\int_0^1 \left[L_{n-1}^1(z \ln t)-n\right]\left({{t^M-1} \over
{t-1}}\right)dt.  \eqno(6.2)$$

{\it Proof}.  We use relation (0.2) together with an integral representation for
the polygamma function,
$${\cal S}_n(M,z)=\sum_{m=2}^n {n \choose m} z^m {{(-1)^{m-1}} \over {(m-1)!}} \left[\psi^{(m-1)}(M+1)-\psi^{(m-1)} (1) \right ]$$
$$=\sum_{m=2}^n {n \choose m} z^m {{(-1)^{m-1}} \over {(m-1)!}} 
\int_0^1 \left({{t^M-1} \over {t-1}}\right) \ln^{m-1} t ~dt. \eqno(6.3)$$
Interchanging the finite summation with the integration and applying the
power series definition of $L_{n-1}^1$, we obtain the Proposition.

{\it Remarks}.  In the limit as $M \to \infty$ in Eq. (6.2), we have the representation
$$\lim_{M \to \infty} {\cal S}_n(M,z) =-z\int_0^1 \left[L_{n-1}^1(z \ln t)-n\right]
{{dt} \over {t-1}}.  \eqno(6.4)$$
This is the same limit in which $H_M^{(m)} \to \zeta(m)$.  In particular, we
put 
$$S_1(n) = \lim_{M \to \infty} \left[{\cal S}_n(M,-1)-{\cal S}_n(M,-1/2)\right].
\eqno(6.5)$$
With a change of variable, we obtain agreement with the integral representation
of $S_1(n)$ of Ref. \cite{coffeympag} (p. 216).  In particular, this alternating binomial sum provides the apparently dominant contribution to the Li/Keiper constants
$\lambda_n$ of the Li criterion for the Riemann hypothesis \cite{coffeympag}.  
This sum has been shown to be $O(n \ln n)$, describing the significant cancellation
within the summand.

{\bf Proposition 10}.  We have
$${\cal S}_n(M,z)=(1+z)^n+\sum_{j=2}^M \left(1+{z \over j}\right)^n-n zH_M-M.
\eqno(6.6)$$

{\it Proof}.
By using the recursion relation of generalized harmonic numbers, we have from the
definition (6.1)
$${\cal S}_n(M,z)={\cal S}_n(M-1,z)-1-{n \over M}z+\left(1+{z \over M}\right)^n.
\eqno(6.7)$$
This recursion may also be found from the integral representation of Proposition 9.
We form the difference
$${\cal S}_n(M,z)-{\cal S}_n(M-1,z) =z\int_0^1 \left[L_{n-1}^1(z \ln t)-n\right] t^{M-1}dt.  \eqno(6.8)$$
We then integrate by parts,
$${\cal S}_n(M,z)-{\cal S}_n(M-1,z) =-\int_0^1 \left({d \over {dt}}L_n(z \ln t)
\right) t^M dt -{{zn} \over M}$$
$$=M\int_0^1 L_n(z \ln t) t^{M-1} dt -L_n(0)-{{zn} \over M}$$
$$=\left(1+ {z \over M}\right)^n -{{zn} \over M}-1.  \eqno(6.9)$$ 
In the last step, we may put $t=\exp(-v)$, and use an extension of the Laplace
transform of the Laguerre polynomial (e.g., formula 7.414.2 of \cite{grad}, with
$\lambda=-z$ and $\mu=0$).

Since we have the initial value $S_n(1,z)=-1-nz+(1+z)^n$, solution of the
recursion (6.7) gives the Proposition.

An alternative proof of Proposition 10 is now obvious.  For from binomial expansion
we obtain
$$\sum_{j=1}^M \left(1+{z \over j}\right)^n=\sum_{j=1}^M \sum_{\ell=0}^n {n \choose
\ell} {z^\ell \over j^\ell}
=\sum_{\ell=0}^n {n \choose \ell} z^\ell H_M^{(\ell)}$$
$$=\sum_{\ell=2}^n {n \choose \ell} z^\ell H_M^{(\ell)}+M+nzH_M.  \eqno(6.10)$$

It is possible to expand the factor $(t-1)^{-1}$ of Eqs. (6.2) and (6.4) as a
geometric series, thereby giving another series representation of the sum
${\cal S}_n$, and providing a third proof of Proposition 9.  
We have
{\newline \bf Corollary 3}.  We have
$${\cal S}_n(M,z)=-nz H_M-M+\sum_{j=1}^\infty \left[\left(1+{z \over j}\right)^n-
\left(1+{z \over {M+j}}\right)^n\right].  \eqno(6.11)$$
A simple shift of summation index here recovers the relation (6.6).

In order to obtain the result (6.11), we write
$${\cal S}_n(M,z)=-z\sum_{j=0}^\infty \int_0^\infty [L_{n-1}^1(-z v)-n][e^{-(M+j+1)v}
-e^{-(j+1)v}] dv.  \eqno(6.12)$$
We then use for Re $k >0$,
$$\int_0^\infty L_{n-1}^1(-zv)e^{-kv}dv={1 \over z}\left[\left(1+{z \over k}\right)^n
-1\right].  \eqno(6.13)$$

{\it Remarks}.  Corollary 3 must be equivalent to using the partial fractions
form of $H_M^{(m)}$ coming from that for the polygamma function, in (6.1), and then 
interchanging summations. 

By manipulating ${\cal S}_n(M,z)$ we may obtain related sums.  For instance,
by integrating Eq. (6.1), using Proposition 8, and interchanging integrations, 
we have
$$\sum_{m=2}^n {n \choose m} {{H_M^{(m)}} \over {m+1}} u^m
=u\int_0^1 \left[{{L_{n-1}^2(u \ln t)} \over {n+1}}-{n \over 2} \right]
\left({{t^M-1} \over {t-1}}\right)dt$$
$$={1 \over u}\sum_{j=2}^n {j \over {(n+1)}}\left[\left(1+{u \over j}\right)^{n+1}
-1\right]-M -{n \over 2}u.  \eqno(6.14)$$

From Proposition 10 we obtain the combination
$${\cal S}_n(M,-1)-{\cal S}_n(M,-1/2)=\sum_{j=1}^M \left[\left(1-{1 \over j}
\right)^n-\left(1-{1 \over {2j}}\right)^n\right]+{n \over 2}H_M$$
$$=\sum_{j=1}^M \left[{n \over 2}{1 \over j}+\left(1-{1 \over j}
\right)^n-\left(1-{1 \over {2j}}\right)^n\right].  \eqno(6.15)$$
The latter expression is the $M < \infty$ form of Eq. (19) of Ref. 
\cite{coffeympag}.

Plainly, the sums of this section may be generalized to those such as
$${\cal S}_n(M,p,q,r,z) \equiv \sum_{m=2}^n m^p{n \choose m}^r [H_M^{(m)}]^q z^m.  \eqno(6.16)$$

\bigskip
\centerline{\bf Acknowledgement}
This work was supported in part by Air Force contract number FA8750-06-1-0001.

\pagebreak

\end{document}